\documentclass{pasj01}
\usepackage{lineno}
\usepackage[usenames,dvipsnames]{xcolor}
\Received{$\langle$reception date$\rangle$}
\Accepted{$\langle$acception date$\rangle$}
\Published{$\langle$publication date$\rangle$}

\newcommand\oiii{[O~{\sc iii}]$_{88}$~}
\newcommand\nii{[N~{\sc ii}]$_{205}$~}
\usepackage{mathpazo}
\usepackage[T1]{fontenc}

\begin{document}

\title{
\LETTERLABEL 
Detection of nitrogen and oxygen in a galaxy at the end of reionization}
\author{Ken-ichi \textsc{Tadaki}\altaffilmark{1,2,*}}%
\author{Akiyoshi \textsc{Tsujita}\altaffilmark{3}}
\author{Yoichi \textsc{Tamura}\altaffilmark{4}}
\author{Kotaro \textsc{Kohno}\altaffilmark{3,5}}
\author{Bunyo \textsc{Hatsukade}\altaffilmark{3}}
\author{Daisuke \textsc{Iono}\altaffilmark{1,2}}
\author{Minju M. \textsc{Lee}\altaffilmark{6,7}}
\author{Yuichi \textsc{Matsuda}\altaffilmark{1,2}}
\author{Tomonari \textsc{Michiyama}\altaffilmark{1,8}}
\author{Tohru \textsc{Nagao}\altaffilmark{9}}
\author{Kouichiro \textsc{Nakanishi}\altaffilmark{1,2}}
\author{Yuri \textsc{Nishimura}\altaffilmark{1,3}}
\author{Toshiki \textsc{Saito}\altaffilmark{1,10}}
\author{Hideki \textsc{Umehata}\altaffilmark{3,11}}
\author{Jorge \textsc{Zavala}\altaffilmark{1}}
\altaffiltext{1}{National Astoronomical Observatory of Japan, 
 2-21-1 Osawa, Mitaka, Tokyo 181-8588, Japan }
 \email{tadaki.ken@nao.ac.jp}
\altaffiltext{2}{Department of Astronomical Science, SOKENDAI (The Graduate University for Advanced Studies), Mitaka, Tokyo 181-8588, Japan}
\altaffiltext{3}{Institute of Astronomy, Graduate School of Science, The University of Tokyo, 2-21-1 Osawa, Mitaka, Tokyo 181-0015, Japan}
\altaffiltext{4}{Division of Particle and Astrophysical Science, Nagoya University, Furocho, Chikusa, Nagoya 464-8602, Japan}
\altaffiltext{5}{Research Center for the Early Universe, Graduate School of Science, The University of Tokyo, 7-3-1 Hongo, Bunkyo-ku, Tokyo 113-0033, Japan}
\altaffiltext{6}{Cosmic Dawn Center (DAWN), Jagtvej 128, DK-2200 Copenhagen N, Denmark}
\altaffiltext{7}{DTU-Space, Technical University of Denmark, Elektrovej 327, DK2800 Kgs. Lyngby, Denmark}
\altaffiltext{8}{Department of Earth and Space Science, Osaka University, 1-1 Machikaneyama, Toyonaka, Osaka 560-0043, Japan}
\altaffiltext{9}{Research Center for Space and Cosmic Evolution, Ehime University, Matsuyama, Ehime 790-8577, Japan}
\altaffiltext{10}{Department of Physics, General Studies, College of Engineering, Nihon University, 1 Nakagawara, Tokusada, Tamuramachi, Koriyama, Fukushima 963-8642, Japan}
\altaffiltext{11}{Institute for Cosmic Ray Research, The University of Tokyo, 5-1-5 Kashiwanoha, Kashiwa, Chiba 277-8582, Japan}

\maketitle

\begin{abstract}
We present observations of [N~{\sc ii}] 205 $\mu$m, [O~{\sc iii}] 88 $\mu$m and dust emission in a strongly-lensed, submillimeter galaxy (SMG) at $z=6.0$, G09.83808, with the Atacama Large Millimeter/submillimeter Array (ALMA).
Both [N~{\sc ii}] and [O~{\sc iii}] line emissions are detected at $>12\sigma$ in the 0.8$''$-resolution maps.
Lens modeling indicates that the spatial distribution of the dust continuum emission is well characterized by a compact disk with an effective radius of 0.64$\pm$0.02 kpc and a high infrared surface brightness of $\Sigma_\mathrm{IR}=(1.8\pm0.3)\times10^{12}~L_\odot$ kpc$^{-2}$.
This result supports that G09.83808 is the progenitors of compact quiescent galaxies at $z\sim4$, where the majority of its stars are expected to be formed through a strong and short burst of star formation.
G09.83808 and other lensed SMGs show a decreasing trend of the [N~{\sc ii}] line to infrared luminosity ratio with increasing continuum flux density ratio between 63 $\mu$m and 158 $\mu$m, as seen in local luminous infrared galaxies (LIRGs).
The decreasing trend can be reproduced by photoionization models with increasing ionization parameters.
Furthermore, by combining the [N~{\sc ii}]/[O~{\sc iii}] luminosity ratio with far-infrared continuum flux density ratio in G09.83808, we infer that the gas phase metallicity is already $Z\approx 0.5-0.7~Z_\odot$.
G09.83808 is likely one of the earliest galaxies that has been chemically enriched at the end of reionization.
\end{abstract}


\section{Introduction}

The most massive galaxies form earliest in the Universe, known as downsizing of galaxy formation (e.g., \cite{1996AJ....112..839C}, \cite{2010MNRAS.404.1775T}).
This naturally motivates us to explore massive mature galaxies at the highest redshift.
The current record redshift of spectroscopically confirmed quiescent galaxies (QGs) is $z=4.01$ \citep{2019ApJ...885L..34T} and many QGs have been identified at $z=3-4$ (e.g., \cite{2018A&A...618A..85S}).
They are extremely compact in the rest-frame optical with an effective radius of less than 1 kpc (e.g., \cite{2018ApJ...867....1K}).
The compact stellar distribution could be related to bursty star formation histories with star formation rate (SFR) of several hundreds $M_\odot$yr$^{-1}$ in the center, as implied by near-infrared spectroscopic studies (e.g., \cite{2017Natur.544...71G}, \cite{2020ApJ...889...93V}).
These findings suggest that starburst galaxies at $z=5-7$ such as bright submillimeter galaxies (SMGs) are good candidates for the progenitors of massive QGs at $z=3-4$ (see also \cite{2014ApJ...782...68T, 2015ApJ...810..133I}).

For understanding how the most massive galaxies grow in such an early universe, we study the star-forming activities and the physical conditions in the interstellar medium (ISM) of a strongly-lensed SMG at $z=6.027$, G09.83808 \citep{2018NatAs...2...56Z}.
G09.83808 is one of three bright SMGs so far discovered at $z>6$, and is a more common populations of starburst galaxies with the intrinsic 870 $\mu$m flux density of $S_\mathrm{870,intr}\sim4$ mJy, compared to the other two extreme ones with $S_\mathrm{870,intr}>10$ mJy \citep{2013Natur.496..329R, 2018Natur.553...51M}.
In this work, we focus on the far-infrared fine structure lines of nitrogen \nii and oxygen [O~{\sc iii}]$_{88}$.
Nitrogen line emission is especially important for understanding the chemical evolution of galaxies because nitrogen is mainly produced from carbon and oxygen already present in stars through the CNO cycle, referred to as secondary element (e.g., \cite{2020ApJ...900..179K}).
Nitrogen is formed in intermediate-mass stars which are longer-lived than massive stars, inducing a time-delay.
Thus, an enhanced ratio between \nii and \oiii luminosity implies that galaxies experienced many cycles of star formation.
Recent ALMA observations have detected nitrogen lines (\nii or [N~{\sc ii}]$_{122}$) in galaxies at $z\sim5$ (e.g., \cite{2019ApJ...882..168P}, \cite{2020MNRAS.494.4090C}). 
But galaxies where both nitrogen and oxygen lines are detected are limited to $z<5$ \citep{2019A&A...631A.167D,2019ApJ...876....1T}, except for bright quasars \citep{2019ApJ...881...63N, 2020ApJ...900..131L}.
For pushing studies of ISM in starburst galaxies to higher redshift,
we observe the \oiii line emission ($\nu_\mathrm{obs}$=482.9 GHz) and \nii ($\nu_\mathrm{obs}$=207.9 GHz), as well as the 0.6 mm and 1.5 mm continuum emission, in G09.83808.

\begin{figure*}[!t]
\begin{center}
\includegraphics[scale=1.1]{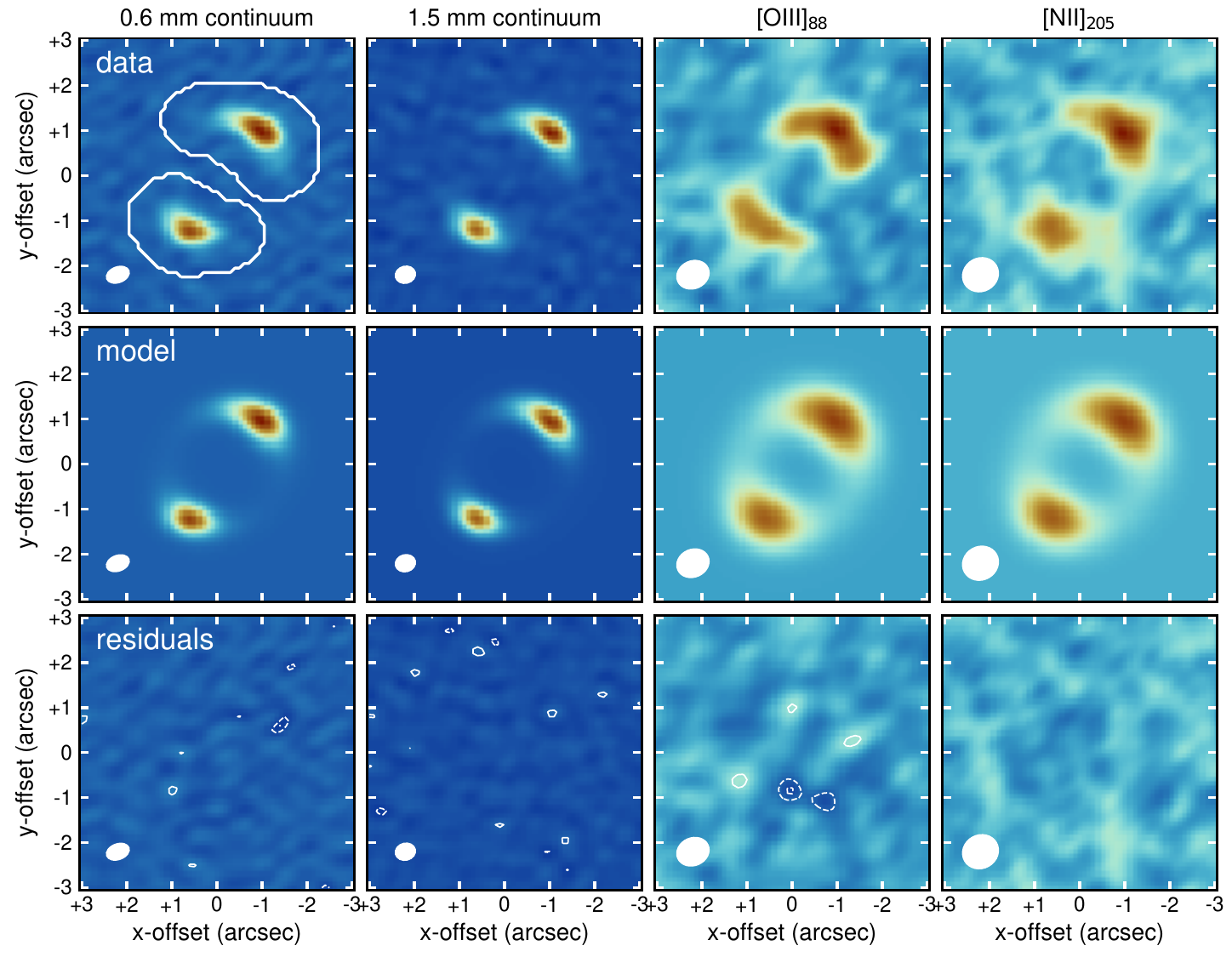}
\end{center}
\caption{
Top: from left to right, the ALMA maps of 0.6 mm, 1.5 mm continuum, [O~{\sc iii}], and \nii line emission are displayed.
The mask region is shown in the top left panel.
Middle and bottom panels show the best-fit models produced by {\tt GLAFIC} and the residuals, respectively.
White dashed and solid contours show the -4$\sigma$, -3$\sigma$ and +3$\sigma$, +4$\sigma$ levels in the residual maps.
We use a scientific color scale from \citet{2020NatCo..11.5444C}.
}
\label{fig;map}
\end{figure*}

\section{Observations}

ALMA observations were executed on 2019 December (Band-8) and 2019 October--2020 January (Band-5). 
On-source time was 2.4 h and 1.6 h, respectively. 
The maximum recovery scale is 5\farcs4 and 6\farcs2, respectively. 
The data were calibrated in the standard manner using {\tt CASA} \citep{2007ASPC..376..127M}.
We first construct a clean mask with the 0.6 mm continuum data by applying {\tt CASA/AUTO-MULTITHRESH} \citep{2020PASP..132b4505K} and Briggs weighting with robust=+0.5.
We use this mask for all imaging of continuum and line emission. 
Then, we clean the emission down to the 1.5$\sigma$ level to create continuum and 100 km s$^{-1}$ channel maps with robust=$-$0.5 and robust=+2.0, respectively.
Figure \ref{fig;map} shows the ALMA maps of the 0.6 mm, 1.5 mm continuum, \oiii and \nii line emission in G09.83808, where two arcs of counter-images are evident. 
The line emission is integrated over the velocity range of $-$350 km s$^{-1}$ to +150 km s$^{-1}$.
The beam size is 0\farcs56$\times$0\farcs38 for the 0.6 mm continuum, 0\farcs48$\times$0\farcs41 for the 1.5 mm continuum, 0\farcs76$\times$0\farcs64 for the \oiii, and 0\farcs84$\times$0\farcs77 for the \nii line map.
The peak flux densities and noise levels are 7.46$\pm$0.17 mJy beam$^{-1}$ (44$\sigma$) for the 0.6 mm continuum, 2.10$\pm$0.03 mJy beam$^{-1}$ (75$\sigma$) for the 1.5 mm continuum, 1.59$\pm$0.12 Jy km s$^{-1}$ beam$^{-1}$ (14$\sigma$) for the \oiii, and 0.38$\pm$0.03 Jy km s$^{-1}$ beam$^{-1}$ (12$\sigma$) for the \nii line map.
The total flux densities and line fluxes in the mask region are 38.69$\pm$1.13 mJy for the 0.6 mm continuum,  9.91$\pm$0.19 mJy for the 1.5 mm continuum, 8.13$\pm$0.50 Jy km s$^{-1}$ for the \oiii, and 1.48$\pm$0.12 Jy km s$^{-1}$ for the \nii line map.
The uncertainties are calculated as 1$\sigma\times\sqrt{(N_\mathrm{mask}/N_\mathrm{beam})}$ where $N_\mathrm{mask}$ and $N_\mathrm{beam}$ is the areas of the mask region and the clean beam, respectively.

\section{Analysis and results}
\subsection{Gravitational lens modeling}
\label{sec;lens_modeling}

Strong gravitational lensing produces multiple images of a background source, G09.83808 at $z=6.027$, in the ALMA maps.
A big advantage of submillimeter observations is that the flux contribution of a foreground source, a massive quiescent galaxy at $z = 0.776$ \citep{2017MNRAS.472.2028F}, is negligible in this wavelength unlike optical and near-infrared observations.
For mass models of foreground (lens) source, we assume a singular isothermal ellipsoid with five parameters ($xy$-coordinates, ellipticity $e$, position angle measured counterclockwise from North $\theta_e$, velocity dispersion $\sigma_v$) and external perturbation with two parameters (tidal shear $\gamma$ and position angle $\theta_\gamma$) in a similar way as in \citet{2015PASJ...67...72T}.
The background source is assumed to have an exponential disk with a S$\acute{\mathrm{e}}$rsic index of $n=1$, characterized by six parameters ($xy$-coordinates, flux, effective radius $R_\mathrm{eff}$, major-to-minor axis ratio $q$ and position angle $\theta_q$), for both the continuum and line emissions.

First, we determine the parameters of the foreground source by using the 1.5 mm continuum image, where the spatial resolution and signal-to-noise ratio are better than those of other images.
We then use {\tt GLAFIC2} software \citep{2010PASJ...62.1017O} to optimize the mass model of the foreground source.
Only the clean mask region is used for $\chi^2$ minimization.
To estimate the uncertainties of the best-fit parameters, we add a $1\sigma$ noise map convolved by a dirty beam to the clean image and repeat to fit the noise added images.
The best-fit parameters are $e=0.11^{+0.02}_{-0.08}$, $\theta_e=78^{+33}_{-3}$ deg, $\sigma_v=257.9^{+0.3}_{-2.0}$ km s$^{-1}$ for an isothermal ellipsoid and $\gamma=4.1^{+2.8}_{-0.0}\times10^{-2}$ and $\theta_\gamma=47^{+11}_{-3}$ deg for an external shear.
The uncertainties are based on the 16th and 84th percentile of 500 MonteCarlo runs.
The derived position of the foreground source is nicely consistent with the position in a deep $z$-band image (0\farcs7 seeing) from the second public data release of the Hyper Suprime-Cam in Subaru Strategic Program \citep{2019PASJ...71..114A}.

We also obtain the central position and the shape of the 1.5 mm continuum emission for the background source.
The spatial distribution is well characterized by an exponential disk with $q=0.93^{+0.02}_{-0.08}$ and $\theta_q=108^{+13}_{-22}$ deg.
Even if $n$ is a free parameter, the best-fit value is $n=1.17^{+0.13}_{-0.10}$, supporting that the dust continuum emission has an exponential profile \citep{2016ApJ...833..103H, 2018ApJ...861....7F}.
The total magnification factor, given by a ratio between flux densities in the image and source plane, is $\mu=8.38^{+0.74}_{-0.27}$, which is consistent with the previous result from 0\farcs1-resolution observations of 870 $\mu$m continuum emission ($\mu=9.3\pm1.0$; \cite{2018NatAs...2...56Z}).

Next, we measure the intrinsic sizes of the continuum and line emissions for the background source by fixing the other parameters to the values obtained above.
The effective radii are 0\farcs112$^{+0.002}_{-0.002}$ for the 1.5 mm continuum, 0\farcs117$^{+0.004}_{-0.003}$ for the 0.6 mm continuum, 0\farcs21$^{+0.08}_{-0.08}$ for the \oiii and 0\farcs20$^{+0.04}_{-0.04}$ for the \nii line emission.
The dust emissions at different wavelength have a similar size and both are more compact than the ionized gas emissions.


Figure \ref{fig;map} shows the best-fit model and residuals for each emission.
The residual image of \oiii emission shows three $+3\sigma$ peaks in the edges of the two arcs, corresponding to the direction of the minor axis of the disk component in the source plane.
Even if the position angle is not fixed, the $3\sigma$ residuals still remain.
Given that the emission peak is detected at 12$\sigma$,
the $3\sigma$ deviation from an exponential disk may suggest that G09.83808 has subcomponents (clumps or small satellite galaxies) of ionized gas.
We require deeper and higher-resolution observations to confirm the existence of these components.

\begin{figure}[!t]
\begin{center}
\includegraphics[scale=1.0]{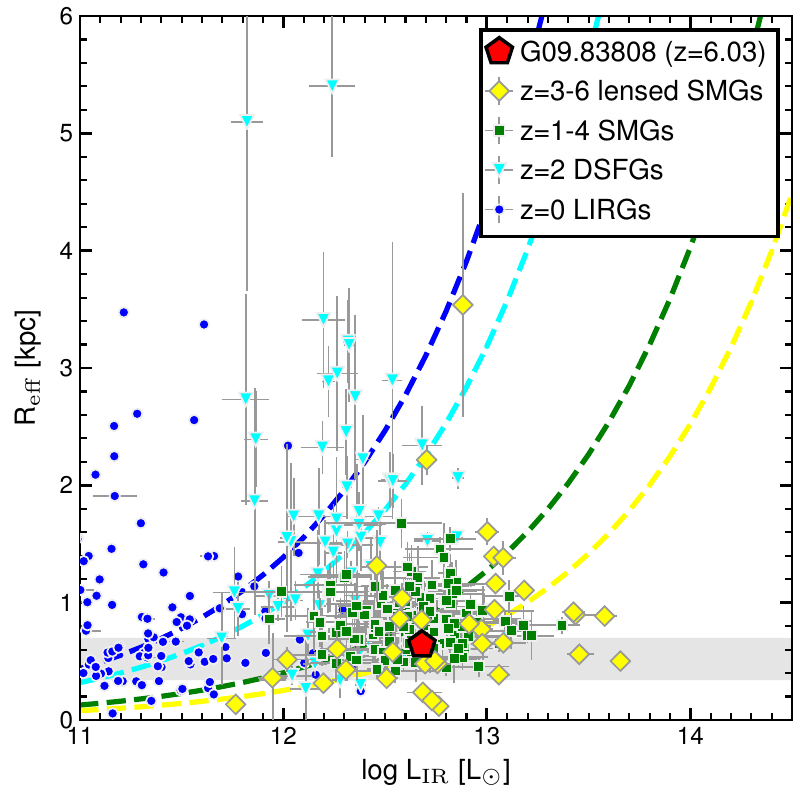}
\end{center}
\caption{
Intrinsic infrared luminosities versus circularized effective radii for G09.83808 at $z=6$, strongly-lensed SMGs (\cite{2016ApJ...826..112S}), SMGs at $z=1-4$ (\cite{2019MNRAS.490.4956G, 2020MNRAS.494.3828D}), DSFGs at $z=2$ (\cite{2020ApJ...901...74T}) and LIRGs at $z=0$ (\cite{2016A&A...591A.136L}). 
A shaded region shows the effective radius of the rest-frame optical emission in massive quiescent galaxies at $z\sim4$ \citep{2018ApJ...867....1K}. 
Dashed lines correspond to the median values of the infrared surface densities for each sample.
}
\label{fig;LIR_size}
\end{figure}

\subsection{Dust SED modeling}
\label{sec;sed}

Combining the new ALMA continuum data with photometry from Herschel, JCMT, and LMT observations \citep{2016ApJ...832...78I, 2017MNRAS.472.2028F, 2018NatAs...2...56Z}, we constrain the spectral energy distribution (SED) of dust emission to estimate the total infrared luminosity, $L_\mathrm{IR}$. 
We assume flux calibration uncertainties of 5\% and 10\% in ALMA Band-5 and Band-8 observations.
We fit a modified blackbody radiation model, characterized by dust temperature $T_\mathrm{dust}$ and an emissivity index $\beta$ \citep{2012ApJ...761..140C}, to the observed SEDs by using the {\tt CIGALE} code \citep{2019A&A...622A.103B}. 
We fix the wavelength where the optical depth is unity to 150 $\mu$m and the power law slope to 2.0.
The best-fit model gives the total infrared luminosity between 8 and 1000 $\mu$m of $L_\mathrm{IR}=(4.6^{+0.6}_{-0.7})\times10^{12}~L_\odot$ after correction of magnification effect, $T_\mathrm{dust}=51\pm4$ K, and $\beta$=$2.5^{+0.3}_{-0.2}$.
As the 0.6 mm continuum emission corresponds to the peak of the dust SED, its spatial distribution probes where stars are intensively formed.
We find G09.83808 to have an infrared surface density of $\Sigma_\mathrm{IR}=(1.8\pm0.3)\times10^{12}~L_\odot$ kpc$^{-2}$ within the circularized effective radius, $R_\mathrm{eff, 0.6 mm}$=0.64$\pm$0.02 kpc.

We also estimate the total infrared luminosities for other strongly lensed SMGs, where the 870 $\mu$m continuum sizes and magnification are measured \citep{2013ApJ...767...88W, 2016ApJ...822...80S, 2016ApJ...826..112S} in the same way as in G09.83808.
Figure \ref{fig;LIR_size} shows the total infrared luminosity and circularized effective radius of dust continuum emission for four different galaxy populations: 1) strongly-lensed SMGs at $z=3-6$ including G09.83808, 2) SMGs at $z=1-4$ \citep{2019MNRAS.490.4956G, 2020MNRAS.494.3828D}, 3) massive dusty star-forming galaxies at $z=2$ (DSFGs; \cite{2020ApJ...901...74T}), and 4) LIRGs at $z=0$ \citep{2016A&A...591A.136L}.
The four populations occupy different regions in the $L_\mathrm{IR}-R_\mathrm{eff}$ plane.
For any combination of the two populations, a KS test shows that the probability that they are drawn from the same distribution is less than 1\%.
The median values of the infrared surface densities are $2.5\times10^{12}~L_\odot$ kpc$^{-2}$ for lensed SMGs, $1.0\times10^{12}~L_\odot$ kpc$^{-2}$ for SMGs, $1.6\times10^{11}~L_\odot$ kpc$^{-2}$ for DSFGs, and $0.8\times10^{11}~L_\odot$ kpc$^{-2}$ for LIRGs.
Thus, lensed SMGs and SMGs are undergoing intense starburst with a higher infrared surface density, compared to LIRGs and DSFGs.
The small difference between lensed SMGs and SMGs may be due to the different redshift range (i.e. emissions at different rest-frame wavelengths) and/or a large magnification near caustics of gravitational lenses in extremely bright objects with $S_\mathrm{1.4~mm}>$20 mJy \citep{2013ApJ...767...88W}.
The effective radii of lensed SMGs are comparable to those of massive QGs at $z\sim4$ ($0.52\pm0.18$ kpc in the rest-frame optical; \cite{2018ApJ...867....1K}).
The intense starburst in the central compact region supports an evolutionary link between G09.83808 at $z=6$ and massive QGs at $z\sim4$.

\begin{figure*}[!t]
\begin{center}
\includegraphics[scale=1.0]{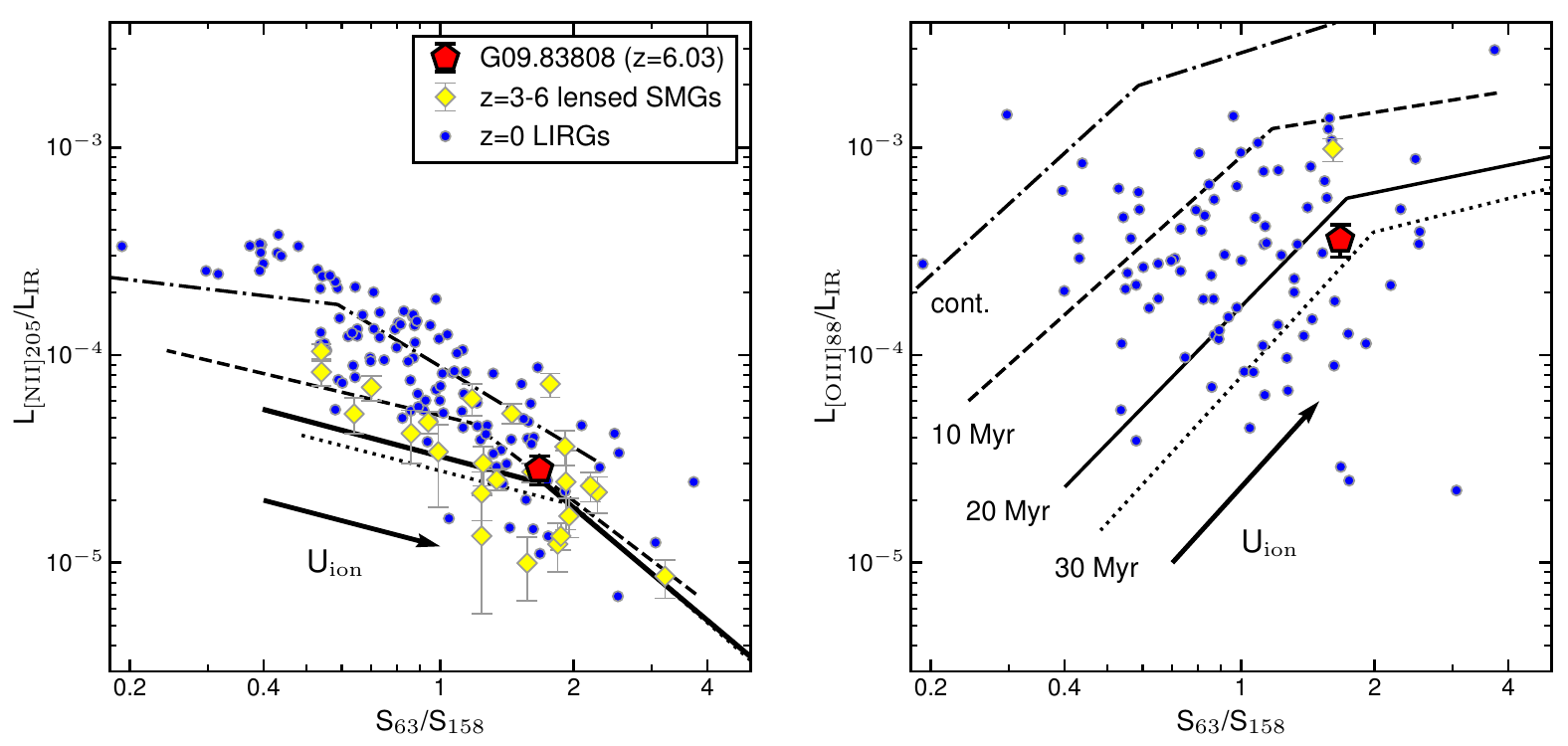}
\end{center}
\caption{
Left: \nii line to infrared luminosity ratio as a function of $S_{63}/S_{158}$ continuum ratio for local LIRGs and strongly-lensed SMGs.
Right: \oiii 88 line to infrared luminosity ratios with the results from single starburst models with an age of 10, 20, and 30 Myr.
A dashed dotted line show continuous star formation models with an age of 20 Myr.
The ionization parameter varies from $U_\mathrm{ion}=10^{-4}$ to $U_\mathrm{ion}=10^{-2}$.
}
\label{fig;line_IR_ratio}
\end{figure*}

\section{Far-infrared line properties}
\label{sec;fir}

UV photons from massive stars ionize the surrounding gas and at the same time heat the dust.
Then, thermal radiation of the dust can be observed in the infrared.
A combination of fine structure lines such as [N~{\sc ii}]$_{205}$ and [O~{\sc iii}]$_{88}$ with infrared continuum emission can therefore provide with information about physical properties of the ISM and ionizing sources (e.g., metallicity, gas density and ionization parameter) in galaxies.
We compare the line-to-infrared luminosity ratio, $L_\mathrm{[NII]205}/L_\mathrm{IR}$ and $L_\mathrm{[OIII]88}/L_\mathrm{IR}$, in G09.83808 and other lensed SMGs at $z=3-6$ with those in local LIRGs.
Although the spatial distributions of ionized gas and dust are different, we use the galaxy-integrated properties for straightforward comparison.

\subsection{Nitrogen line emission}
\label{sec;nii_ir}

$Herschel$ observations of [N~{\sc ii}]$_{205}$ line in LIRGs at $z\sim0$ show that $L_\mathrm{[NII]205}/L_\mathrm{IR}$ is anticorrelated with the continuum flux density ratio between 63 $\mu$m and 158 $\mu$m, $S_{63}/S_{158}$ \citep{2017ApJ...846...32D,2017ApJS..230....1L}.
For lensed SMGs, we derive $S_{63}/S_{158}$ from the best-fit SED (section \ref{sec;sed}).
In both galaxy populations, there is a decreasing trend of $L_\mathrm{[NII]205}/L_\mathrm{IR}$ with increasing $S_{63}/S_{158}$ (Figure \ref{fig;line_IR_ratio}).
The trends could be related to a variation in ionization parameter, defined as $U_\mathrm{ion}=\phi(H)/c/n_H$ where $\phi(H)$ is the flux of hydrogen-ionizing photons, $n_H$ is the hydrogen density at the illuminated face of a cloud, and $c$ is the light speed.
We use photoionization code {\tt Cloudy~v17.01} \citep{2017RMxAA..53..385F} to compare the observations with models with different ionization parameters. 
We generate the input spectra of a single age starburst model with 20 Myr by using the Binary Population and Spectral Synthesis ({\tt BPASS v2.0}) code \citep{2016MNRAS.462.3302E}. 
The initial gas density at illuminated face is fixed to be $n_H$=50 cm$^{-3}$, which is the typical value in local LIRGs \citep{2017ApJ...846...32D}.
We adopt solar elemental abundance ratios and gas-phase depletion factors, with taking into account secondary production of nitrogen \citep{2011A&A...526A.149N} and assume a solar metalliciity.
For dust, we assume Orion-type graphite and silicate grains with a size distribution and abundance appropriate for those along the line of sight to the Trapezium stars in Orion.
We stop calculations at the total hydrogen column density of $N(H)=10^{22}$ cm$^{-2}$ to avoid the dust temperature becoming too low.
We here do not intend to determine each parameters from fitting, but aim to interpret the observed trends from comparison with models.

The decreasing trend of $L_\mathrm{[NII]205}/L_\mathrm{IR}$ is successfully reproduced by photoionization models in the range of $U_\mathrm{ion}=10^{-4}-10^{-2}$ (Figure \ref{fig;line_IR_ratio}).
As an ionization parameter becomes larger, the H$^+$ region expands.
But the volume of H$^+$ region does not increase linearly with $U_\mathrm{ion}$ because UV photon in turn is used to heat the dust in the expanded H$^+$ region (\cite{2009ApJ...701.1147A}).
The fraction of UV photon available for ionization becomes smaller while all of its energy is eventually converted into dust emission, resulting in a decrease of $L_\mathrm{[NII]205}/L_\mathrm{IR}$.
On the other hand since the UV photon per dust particle increases, the dust temperature becomes higher, and then $S_{63}/S_{158}$ becomes larger \citep{2009ApJ...701.1147A, 2018MNRAS.473...20R}.
Therefore, in both local and high-redshift galaxies, the decreasing trend can be explained by higher ionization paremters.

\subsection{Oxygen line emission}

Unlike $L_\mathrm{[NII]205}/L_\mathrm{IR}$, the photoionization models predicts increasing $L_\mathrm{[OIII]88}/L_\mathrm{IR}$ with increasing ionization parameter (Figuref \ref{fig;line_IR_ratio}).
This is because only a small fraction ($<$10\%) of oxygen is doubly ionized at low ionization parameter.
Nevertheless, local LIRGs do not show any correlation between $L_\mathrm{[OIII]88}/L_\mathrm{IR}$ and $S_{63}/S_{158}$.
The $L_\mathrm{[OIII]88}/L_\mathrm{IR}$ values of two lensed SMGs (G09.83808 at $z=6.0$ and SPT 0418--47 at $z=4.2$; \cite{2019A&A...631A.167D}) are also consistent with those in local LIRGs.
From comparisons with photoionization models, we find that $L_\mathrm{[OIII]88}/L_\mathrm{IR}$ is very sensitive to a variation in age of star formation, which changes the energy distribution of incident radiation.
The contribution of massive stars to the incident radiation is larger for ages younger than 20 Myr, leading to higher $L_\mathrm{[OIII]88}/L_\mathrm{IR}$.
The trend with changing age is almost orthogonal to the trend with changing ionization parameter.
Therefore, the ionization parameter dependence of $L_\mathrm{[OIII]88}/L_\mathrm{IR}$ quickly disappear due to a small variation in age of star formation.
The impact of age variation on \nii is small because both trends are parallel in the $L_\mathrm{[NII]205}/L_\mathrm{IR}$--$S_{63}/S_{158}$ plane.
Single starburst model with even the younger age ($<10$ Myr) and continuous star formation models predict a much higher $L_\mathrm{[OIII]88}/L_\mathrm{IR}$ of $10^{-3}-10^{-2}$, which is similar to those in local dwarf galaxies \citep{2015A&A...578A..53C}.

We also note that these arguments are based on simple spherical models in which all of [N~{\sc ii}]$_{205}$, [O~{\sc iii}]$_{88}$ and dust emissions are radiated from the same clouds.
If [O~{\sc iii}]$_{88}$ emission comes from a high density gas ($n_H>$1000 cm$^{-3}$) unlike [N~{\sc ii}]$_{205}$, photoionization models with a different column density can reproduce a large variation in $L_\mathrm{[OIII]88}/L_\mathrm{IR}$ ratio \citep{2014ApJ...795..117F}.

\section{Gas-phase metallicity}

In this section, we estimate the gas-phase metallicity by using two measurements of [N~{\sc ii}]$_{205}$/\oiii and $S_{63}/S_{158}$ ratio in G09.83808.
Since \nii and \oiii lines have different critical densities and ionization potential, its ratio depends not only on metallicity but also on gas density and ionization parameter.
In local LIRGs, the gas density is mostly in a narrow range of $n_e$=20--100 cm$^{-3}$ \citep{2017ApJ...846...32D}.
We assume that the $z=6$ galaxy has a line ratio of $\log(L_\mathrm{[NII]122}/L_\mathrm{[NII]205}) = 0.20 \pm 0.18$, which is the median value in LIRGs.
This assumption is consistent with previous observations of SMGs at $z\sim4$ \citep{2019A&A...631A.167D, 2019ApJ...883L..29L}.
The ionization parameter dependence more seriously affects the estimates of metallicities when a [N~{\sc ii}]$_{122}$/\oiii line ratio is used.
[N~{\sc iii}]$_{57}$/\oiii with similar ionization potential is considered to be a better indicator of metallicity \citep{2011A&A...526A.149N, 2017MNRAS.470.1218P}.
However, either of [N~{\sc iii}]$_{57}$ and \oiii lines at $z>6$ is shifted to be in frequency ranges where the atmospheric transmission is low or even zero, except for $z=6.8-6.9$ and $z=7.1-7.3$.
In addition, as [N~{\sc iii}]$_{57}$ emission at $z\sim7$ can be observed with ALMA Band-9 receivers, the required integration time become by a factor of 60 larger than that in Band-7 observations of [N~{\sc ii}]$_{122}$ line at the same limiting flux.
An approach of using a [N~{\sc ii}]$_{122}$/\oiii line ratio therefore has a great advantage for future measurements of metallicity for a large sample once the dependence of the ionization parameter is taken into account.

\begin{figure}[!t]
\begin{center}
\includegraphics[scale=1.0]{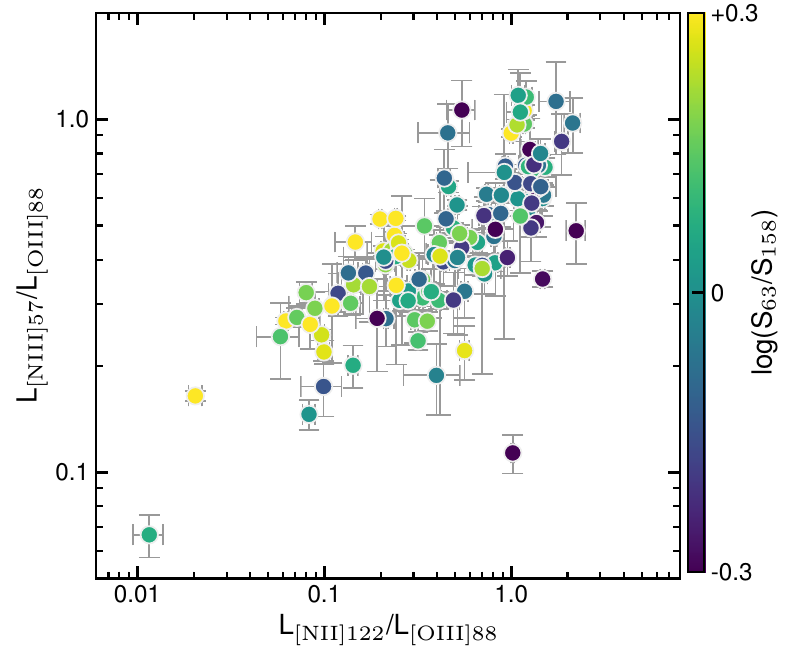}
\end{center}
\caption{
A comparison between $L_\mathrm{[NIII]57}/L_\mathrm{[OIII]88}$ and $L_\mathrm{[NII]122}/L_\mathrm{[OIII]88}$ for galaxies $z=0$ \citep{2015A&A...578A..53C,2016ApJS..226...19F, 2018ApJ...861...94H}.
Color coding shows the $S_{63}/S_{158}$ continuum ratio.
}
\label{fig;NIII57}
\end{figure}

In local galaxies where both lines are detected \citep{2015A&A...578A..53C,2016ApJS..226...19F, 2018ApJ...861...94H}, the [N~{\sc iii}]$_{57}$/\oiii ratio is correlated with the [N~{\sc ii}]$_{122}$/\oiii ratio, though with a large dispersion (Figure \ref{fig;NIII57}).
At a similar [N~{\sc iii}]$_{57}$/\oiii ratio, galaxies with a higher [N~{\sc ii}]$_{122}$/\oiii ratio tend to have a lower $S_{63}/S_{158}$ ratio, corresponding to a lower ionization parameter.
We therefore introduce the scaling relation to predict [N~{\sc iii}]$_{57}$/\oiii ratio as $\log(L_\mathrm{[NIII]57}/L_\mathrm{[OIII]88})_\mathrm{pred}=A+B\log(L_{[NII]205}/L_{[OIII]88})+C\log(S_{63}/S_{158})$.
We determine the coefficients to minimize the difference between the predicted and observed [N~{\sc iii}]$_{57}$/\oiii in local galaxies by ordinary least squares regression.
Thus, we obtain $A$=--0.21$\pm$0.02, $B$=0.45$\pm$0.03 and $C$=0.32$\pm$0.07, with a dispersion of $\pm$0.14 dex in $\log(L_\mathrm{[NIII]57}/L_\mathrm{[OIII]88})$.

\begin{figure}[!t]
\begin{center}
\includegraphics[scale=1.0]{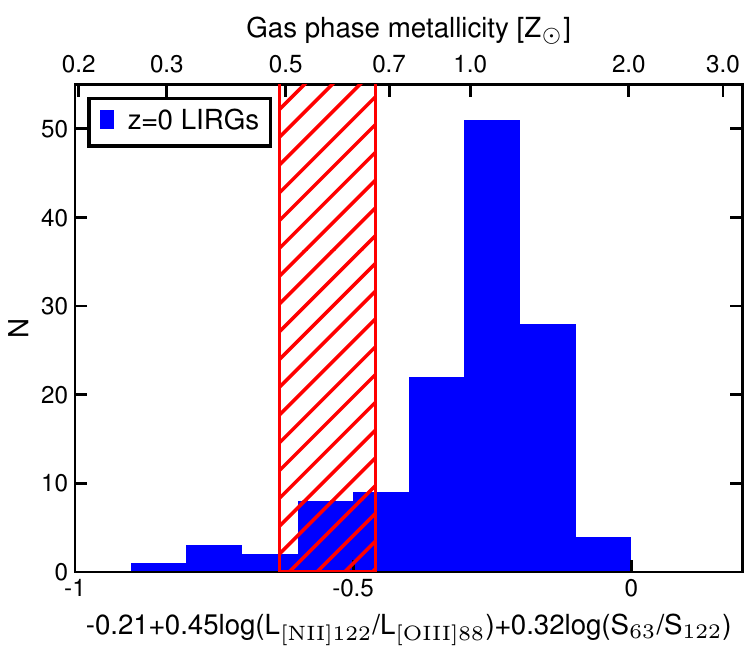}
\end{center}
\caption{
A histogram of [N~{\sc iii}]$_{57}$/\oiii ratio inferred from the scaling relation for local LIRGs (blue; \cite{2017ApJ...846...32D}).
A red hatched region shows the range of G09.83808 at $z=6$, including uncertainties on conversion from [N~{\sc ii}]$_{205}$ to [N~{\sc ii}]$_{122}$.
The top x-axis denotes gas-phase metallicities based on the photoionization models (section \ref{sec;nii_ir}).
}
\label{fig;hist}
\end{figure}

By using the scaling relation, we infer log([N~{\sc iii}]$_{57}$/[O~{\sc iii}]$_{88}$)=--0.55$\pm$0.09 where the uncertainty includes that due to the conversion from [N~{\sc ii}]$_{205}$ to [N~{\sc ii}]$_{122}$ as well as the measurement errors.
This ratio is relatively low compared to local LIRGs (Figure \ref{fig;hist}), but implies $Z=0.5-0.7~Z_\odot$ according to the photoionization models (section \ref{sec;nii_ir}).
Our result is consistent with previous studies, where it is claimed that SMGs at $z=3-4$ are chemically evolved with nearly solar metallicity (e.g., \cite{2018MNRAS.473...20R,2019ApJ...876....1T, 2019A&A...631A.167D}).
Numerical simulations also predict $Z\sim0.5~Z_\odot$ at $z=6$ in the stellar mass range of $\log(M_\star/M_\odot)=10-10.5$ \citep{2019MNRAS.484.5587T}.
High-resolution 3--4 $\mu$m observations with the James Webb Space Telescope (JWST) will allow us to obtain the stellar mass of strongly-lensed galaxies at $z=4-6$ in separate from a foreground object. 
Therefore, the ALMA--JWST synergetic observations will allow us to probe the massive end of stellar mass-metallicity relation for galaxies at $z=6$, which has not been explored so far.

\begin{ack}
We thank the referee for constructive comments that improved the paper. 
We wish to thank Jacqueline Fischer for advice about photoionization modeling with {\tt Cloudy}.
We also would like to thank Tanio D\'{i}az-Santos and Rodrigo Herrera-Camus for kindly providing catalogs of galaxies at $z\sim0$.
This paper makes use of the following ALMA data: ADS/JAO.ALMA\#2019.1.01307.S. ALMA is a partnership of ESO (representing its member states), NSF (USA) and NINS (Japan), together with NRC (Canada), MOST and ASIAA (Taiwan), and KASI (Republic of Korea), in cooperation with the Republic of Chile. The Joint ALMA Observatory is operated by ESO, AUI/NRAO and NAOJ.
We thank the ALMA staff and in particular the EA-ARC staff for their support.
This work was supported by JSPS KAKENHI Grant Numbers 20K14526, 17H06130.
Data analysis was in part carried out on the Multi-wavelength Data Analysis System operated by the Astronomy Data Center (ADC), National Astronomical Observatory of Japan.
\end{ack}

\bibliographystyle{apj}
\bibliography{ref}

\begin{thebibliography}{}
\expandafter\ifx\csname natexlab\endcsname\relax\def\natexlab#1{#1}\fi

\bibitem[{{Abel} {et~al.}(2009){Abel}, {Dudley}, {Fischer}, {Satyapal}, \& {van
  Hoof}}]{2009ApJ...701.1147A}
{Abel}, N.~P., {Dudley}, C., {Fischer}, J., {Satyapal}, S., \& {van Hoof},
  P.~A.~M. 2009, \apj, 701, 1147

\bibitem[{{Aihara} {et~al.}(2019){Aihara}, {AlSayyad}, {Ando}, {Armstrong},
  {Bosch}, {Egami}, {Furusawa}, {Furusawa}, {Goulding}, {Harikane}, {Hikage},
  {Ho}, {Hsieh}, {Huang}, {Ikeda}, {Imanishi}, {Ito}, {Iwata}, {Jaelani},
  {Kakuma}, {Kawana}, {Kikuta}, {Kobayashi}, {Koike}, {Komiyama}, {Li},
  {Liang}, {Lin}, {Luo}, {Lupton}, {Lust}, {MacArthur}, {Matsuoka}, {Mineo},
  {Miyatake}, {Miyazaki}, {More}, {Murata}, {Namiki}, {Nishizawa}, {Oguri},
  {Okabe}, {Okamoto}, {Okura}, {Ono}, {Onodera}, {Onoue}, {Osato}, {Ouchi},
  {Shibuya}, {Strauss}, {Sugiyama}, {Suto}, {Takada}, {Takagi}, {Takata},
  {Takita}, {Tanaka}, {Terai}, {Toba}, {Uchiyama}, {Utsumi}, {Wang}, {Wang}, \&
  {Yamada}}]{2019PASJ...71..114A}
{Aihara}, H., {et~al.} 2019, \pasj, 71, 114

\bibitem[{{Boquien} {et~al.}(2019){Boquien}, {Burgarella}, {Roehlly}, {Buat},
  {Ciesla}, {Corre}, {Inoue}, \& {Salas}}]{2019A&A...622A.103B}
{Boquien}, M., {Burgarella}, D., {Roehlly}, Y., {Buat}, V., {Ciesla}, L., {Corre}, D., {Inoue}, A.~K., \& {Salas}, H. 2019, \aap, 622, A103

\bibitem[{{Casey} {et~al.}(2012){Casey}, {Berta}, {B{\'e}thermin}, {Bock},
  {Bridge}, {Budynkiewicz}, {Burgarella}, {Chapin}, {Chapman}, {Clements},
  {Conley}, {Conselice}, {Cooray}, {Farrah}, {Hatziminaoglou}, {Ivison}, {le
  Floc'h}, {Lutz}, {Magdis}, {Magnelli}, {Oliver}, {Page}, {Pozzi},
  {Rigopoulou}, {Riguccini}, {Roseboom}, {Sanders}, {Scott}, {Seymour},
  {Valtchanov}, {Vieira}, {Viero}, \& {Wardlow}}]{2012ApJ...761..140C}
{Casey}, C.~M., {et~al.} 2012, \apj, 761, 140

\bibitem[{{Cormier} {et~al.}(2015){Cormier}, {Madden}, {Lebouteiller}, {Abel},
  {Hony}, {Galliano}, {R{\'e}my-Ruyer}, {Bigiel}, {Baes}, {Boselli},
  {Chevance}, {Cooray}, {De Looze}, {Doublier}, {Galametz}, {Hughes},
  {Karczewski}, {Lee}, {Lu}, \& {Spinoglio}}]{2015A&A...578A..53C}
{Cormier}, D., {et~al.} 2015, \aap, 578,
  A53

\bibitem[{{Cowie} {et~al.}(1996){Cowie}, {Songaila}, {Hu}, \&
  {Cohen}}]{1996AJ....112..839C}
{Cowie}, L.~L., {Songaila}, A., {Hu}, E.~M., \& {Cohen}, J.~G. 1996, \aj, 112,
  839

\bibitem[{{Crameri} {et~al.}(2020){Crameri}, {Shephard}, \&
  {Heron}}]{2020NatCo..11.5444C}
{Crameri}, F., {Shephard}, G.~E., \& {Heron}, P.~J. 2020, Nature
  Communications, 11, 5444

\bibitem[{{Cunningham} {et~al.}(2020){Cunningham}, {Chapman}, {Aravena}, {De
  Breuck}, {B{\'e}thermin}, {Chen}, {Dong}, {Gonzalez}, {Greve}, {Litke}, {Ma},
  {Malkan}, {Marrone}, {Miller}, {Phadke}, {Reuter}, {Rotermund}, {Spilker},
  {Stark}, {Strandet}, {Vieira}, \& {Wei{\ss}}}]{2020MNRAS.494.4090C}
{Cunningham}, D.~J.~M., {et~al.} 2020, \mnras,
  494, 4090

\bibitem[{{De Breuck} {et~al.}(2019){De Breuck}, {Wei{\ss}}, {B{\'e}thermin},
  {Cunningham}, {Apostolovski}, {Aravena}, {Archipley}, {Chapman}, {Chen},
  {Fu}, {Jarugula}, {Malkan}, {Mangian}, {Phadke}, {Reuter}, {Stacey},
  {Strandet}, {Vieira}, \& {Vishwas}}]{2019A&A...631A.167D}
{De Breuck}, C., {et~al.} 2019, \aap, 631,
  A167

\bibitem[{{D{\'\i}az-Santos} {et~al.}(2017){D{\'\i}az-Santos}, {Armus},
  {Charmandaris}, {Lu}, {Stierwalt}, {Stacey}, {Malhotra}, {van der Werf},
  {Howell}, {Privon}, {Mazzarella}, {Goldsmith}, {Murphy}, {Barcos-Mu{\~n}oz},
  {Linden}, {Inami}, {Larson}, {Evans}, {Appleton}, {Iwasawa}, {Lord},
  {Sanders}, \& {Surace}}]{2017ApJ...846...32D}
{D{\'\i}az-Santos}, T., {et~al.} 2017, \apj,
  846, 32

\bibitem[{{Dudzevi{\v{c}}i{\={u}}t{\.{e}}}
  {et~al.}(2020){Dudzevi{\v{c}}i{\={u}}t{\.{e}}}, {Smail}, {Swinbank}, {Stach},
  {Almaini}, {da Cunha}, {An}, {Arumugam}, {Birkin}, {Blain}, {Chapman},
  {Chen}, {Conselice}, {Coppin}, {Dunlop}, {Farrah}, {Geach}, {Gullberg},
  {Hartley}, {Hodge}, {Ivison}, {Maltby}, {Scott}, {Simpson}, {Simpson},
  {Thomson}, {Walter}, {Wardlow}, {Weiss}, \& {van der
  Werf}}]{2020MNRAS.494.3828D}
{Dudzevi{\v{c}}i{\={u}}t{\.{e}}}, U., {et~al.}
  2020, \mnras, 494, 3828

\bibitem[{{Eldridge} \& {Stanway}(2016)}]{2016MNRAS.462.3302E}
{Eldridge}, J.~J., \& {Stanway}, E.~R. 2016, \mnras, 462, 3302

\bibitem[{{Ferland} {et~al.}(2017){Ferland}, {Chatzikos}, {Guzm{\'a}n},
  {Lykins}, {van Hoof}, {Williams}, {Abel}, {Badnell}, {Keenan}, {Porter}, \&
  {Stancil}}]{2017RMxAA..53..385F}
{Ferland}, G.~J., {et~al.} 2017, RMxAA, 53,
  385

\bibitem[{{Fern{\'a}ndez-Ontiveros} {et~al.}(2016){Fern{\'a}ndez-Ontiveros},
  {Spinoglio}, {Pereira-Santaella}, {Malkan}, {Andreani}, \&
  {Dasyra}}]{2016ApJS..226...19F}
{Fern{\'a}ndez-Ontiveros}, J.~A., {et~al.} 2016, \apjs, 226, 19

\bibitem[{{Fischer} {et~al.}(2014){Fischer}, {Abel}, {Gonz{\'a}lez-Alfonso},
  {Dudley}, {Satyapal}, \& {van Hoof}}]{2014ApJ...795..117F}
{Fischer}, J., {Abel}, N.~P., {Gonz{\'a}lez-Alfonso}, E., {Dudley}, C.~C., {Satyapal}, S., \& {van Hoof}, P.~A.~M. 2014, \apj,
  795, 117

\bibitem[{{Fudamoto} {et~al.}(2017){Fudamoto}, {Ivison}, {Oteo}, {Krips},
  {Zhang}, {Weiss}, {Dannerbauer}, {Omont}, {Chapman}, {Christensen},
  {Arumugam}, {Bertoldi}, {Bremer}, {Clements}, {Dunne}, {Eales}, {Greenslade},
  {Maddox}, {Martinez-Navajas}, {Michalowski}, {P{\'e}rez-Fournon}, {Riechers},
  {Simpson}, {Stalder}, {Valiante}, \& {van der Werf}}]{2017MNRAS.472.2028F}
{Fudamoto}, Y., {et~al.} 2017, \mnras, 472, 2028

\bibitem[{{Fujimoto} {et~al.}(2018){Fujimoto}, {Ouchi}, {Kohno}, {Yamaguchi},
  {Hatsukade}, {Ueda}, {Shibuya}, {Inoue}, {Oogi}, {Toft},
  {G{\'o}mez-Guijarro}, {Wang}, {Espada}, {Nagao}, {Tanaka}, {Ao}, {Umehata},
  {Taniguchi}, {Nakanishi}, {Rujopakarn}, {Ivison}, {Wang}, {Lee}, {Tadaki},
  {Tamura}, \& {Dunlop}}]{2018ApJ...861....7F}
{Fujimoto}, S., {Ouchi}, M., {Kohno}, K., {et~al.} 2018, \apj, 861, 7

\bibitem[{{Glazebrook} {et~al.}(2017){Glazebrook}, {Schreiber}, {Labb{\'e}},
  {Nanayakkara}, {Kacprzak}, {Oesch}, {Papovich}, {Spitler}, {Straatman},
  {Tran}, \& {Yuan}}]{2017Natur.544...71G}
{Glazebrook}, K., {et~al.} 2017, \nat, 544,
  71

\bibitem[{{Gullberg} {et~al.}(2019){Gullberg}, {Smail}, {Swinbank},
  {Dudzevi{\v{c}}i{\={u}}t{\.{e}}}, {Stach}, {Thomson}, {Almaini}, {Chen},
  {Conselice}, {Cooke}, {Farrah}, {Ivison}, {Maltby}, {Micha{\l}owski},
  {Simpson}, {Scott}, {Wardlow}, \& {Weiss}}]{2019MNRAS.490.4956G}
{Gullberg}, B., {et~al.} 2019, \mnras, 490,
  4956

\bibitem[{{Herrera-Camus} {et~al.}(2018){Herrera-Camus}, {Sturm},
  {Graci{\'a}-Carpio}, {Lutz}, {Contursi}, {Veilleux}, {Fischer},
  {Gonz{\'a}lez-Alfonso}, {Poglitsch}, {Tacconi}, {Genzel}, {Maiolino},
  {Sternberg}, {Davies}, \& {Verma}}]{2018ApJ...861...94H}
{Herrera-Camus}, R., {et~al.} 2018, \apj,
  861, 94

\bibitem[{{Hodge} {et~al.}(2016){Hodge}, {Swinbank}, {Simpson}, {Smail},
  {Walter}, {Alexander}, {Bertoldi}, {Biggs}, {Brandt}, {Chapman}, {Chen},
  {Coppin}, {Cox}, {Dannerbauer}, {Edge}, {Greve}, {Ivison}, {Karim},
  {Knudsen}, {Menten}, {Rix}, {Schinnerer}, {Wardlow}, {Weiss}, \& {van der
  Werf}}]{2016ApJ...833..103H}
{Hodge}, J.~A., {et~al.} 2016, \apj, 833,
  103

\bibitem[{{Ikarashi} {et~al.}(2015){Ikarashi}, {Ivison}, {Caputi}, {Aretxaga},
  {Dunlop}, {Hatsukade}, {Hughes}, {Iono}, {Izumi}, {Kawabe}, {Kohno}, {Lagos},
  {Motohara}, {Nakanishi}, {Ohta}, {Tamura}, {Umehata}, {Wilson}, {Yabe}, \&
  {Yun}}]{2015ApJ...810..133I}
{Ikarashi}, S., {et~al.} 2015, \apj, 810, 133

\bibitem[{{Ivison} {et~al.}(2016){Ivison}, {Lewis}, {Weiss}, {Arumugam},
  {Simpson}, {Holland}, {Maddox}, {Dunne}, {Valiante}, {van der Werf}, {Omont},
  {Dannerbauer}, {Smail}, {Bertoldi}, {Bremer}, {Bussmann}, {Cai}, {Clements},
  {Cooray}, {De Zotti}, {Eales}, {Fuller}, {Gonzalez-Nuevo}, {Ibar},
  {Negrello}, {Oteo}, {P{\'e}rez-Fournon}, {Riechers}, {Stevens}, {Swinbank},
  \& {Wardlow}}]{2016ApJ...832...78I}
{Ivison}, R.~J., {et~al.} 2016, \apj, 832, 78

\bibitem[{{Kepley} {et~al.}(2020){Kepley}, {Tsutsumi}, {Brogan}, {Indebetouw},
  {Yoon}, {Mason}, \& {Donovan Meyer}}]{2020PASP..132b4505K}
{Kepley}, A.~A., {Tsutsumi}, T., {Brogan}, C.~L., {Indebetouw}, R., {Yoon}, I., {Mason}, B., \& {Donovan Meyer}, J. 2020, \pasp, 132, 024505

\bibitem[{{Kobayashi} {et~al.}(2020){Kobayashi}, {Karakas}, \&
  {Lugaro}}]{2020ApJ...900..179K}
{Kobayashi}, C., {Karakas}, A.~I., \& {Lugaro}, M. 2020, \apj, 900, 179

\bibitem[{{Kubo} {et~al.}(2018){Kubo}, {Tanaka}, {Yabe}, {Toft}, {Stockmann},
  \& {G{\'o}mez-Guijarro}}]{2018ApJ...867....1K}
{Kubo}, M., {Tanaka}, M., {Yabe}, K., {Toft}, S., {Stockmann}, M., \& {G{\'o}mez-Guijarro}, C. 2018, \apj, 867, 1

\bibitem[{{Lee} {et~al.}(2019){Lee}, {Nagao}, {De Breuck}, {Carniani},
  {Cresci}, {Hatsukade}, {Kawabe}, {Kohno}, {Maiolino}, {Mannucci}, {Marconi},
  {Nakanishi}, {Saito}, {Tamura}, {Troncoso}, {Umehata}, \&
  {Yun}}]{2019ApJ...883L..29L}
{Lee}, M.~M., {et~al.} 2019, \apjl, 883, L29

\bibitem[{{Li} {et~al.}(2020){Li}, {Wang}, {Cox}, {Gao}, {Walter}, {Wagg},
  {Menten}, {Bertoldi}, {Shao}, {Venemans}, {Decarli}, {Riechers}, {Neri},
  {Fan}, {Omont}, \& {Narayanan}}]{2020ApJ...900..131L}
{Li}, J., {et~al.} 2020, \apj, 900, 131

\bibitem[{{Lu} {et~al.}(2017){Lu}, {Zhao}, {D{\'\i}az-Santos}, {Xu}, {Gao},
  {Armus}, {Isaak}, {Mazzarella}, {van der Werf}, {Appleton}, {Charmandaris},
  {Evans}, {Howell}, {Iwasawa}, {Leech}, {Lord}, {Petric}, {Privon}, {Sanders},
  {Schulz}, \& {Surace}}]{2017ApJS..230....1L}
{Lu}, N., {et~al.} 2017, \apjs, 230, 1

\bibitem[{{Lutz} {et~al.}(2016){Lutz}, {Berta}, {Contursi}, {F{\"o}rster
  Schreiber}, {Genzel}, {Graci{\'a}-Carpio}, {Herrera-Camus}, {Netzer},
  {Sturm}, {Tacconi}, {Tadaki}, \& {Veilleux}}]{2016A&A...591A.136L}
{Lutz}, D., {et~al.} 2016, \aap, 591, A136

\bibitem[{{Marrone} {et~al.}(2018){Marrone}, {Spilker}, {Hayward}, {Vieira},
  {Aravena}, {Ashby}, {Bayliss}, {B{\'e}thermin}, {Brodwin}, {Bothwell},
  {Carlstrom}, {Chapman}, {Chen}, {Crawford}, {Cunningham}, {De Breuck},
  {Fassnacht}, {Gonzalez}, {Greve}, {Hezaveh}, {Lacaille}, {Litke}, {Lower},
  {Ma}, {Malkan}, {Miller}, {Morningstar}, {Murphy}, {Narayanan}, {Phadke},
  {Rotermund}, {Sreevani}, {Stalder}, {Stark}, {Strandet}, {Tang}, \&
  {Wei{\ss}}}]{2018Natur.553...51M}
{Marrone}, D.~P., {et~al.} 2018, \nat, 553,
  51

\bibitem[{{McMullin} {et~al.}(2007){McMullin}, {Waters}, {Schiebel}, {Young},
  \& {Golap}}]{2007ASPC..376..127M}
{McMullin}, J.~P., {Waters}, B., {Schiebel}, D., {Young}, W., \& {Golap}, K. 
  2007, in Astronomical Society of the Pacific Conference Series, Vol. 376,
  Astronomical Data Analysis Software and Systems XVI, ed. R.~A. {Shaw},
  F.~{Hill}, \& D.~J. {Bell}, 127

\bibitem[{{Nagao} {et~al.}(2011){Nagao}, {Maiolino}, {Marconi}, \&
  {Matsuhara}}]{2011A&A...526A.149N}
{Nagao}, T., {Maiolino}, R., {Marconi}, A., \& {Matsuhara}, H. 2011, \aap, 526,
  A149

\bibitem[{{Novak} {et~al.}(2019){Novak}, {Ba{\~n}ados}, {Decarli}, {Walter},
  {Venemans}, {Neeleman}, {Farina}, {Mazzucchelli}, {Carilli}, {Fan}, {Rix}, \&
  {Wang}}]{2019ApJ...881...63N}
{Novak}, M., {et~al.} 2019, \apj, 881, 63

\bibitem[{{Oguri}(2010)}]{2010PASJ...62.1017O}
{Oguri}, M. 2010, \pasj, 62, 1017

\bibitem[{{Pavesi} {et~al.}(2019){Pavesi}, {Riechers}, {Faisst}, {Stacey}, \&
  {Capak}}]{2019ApJ...882..168P}
{Pavesi}, R., {Riechers}, D. A., {Faisst}, A. L., {Stacey}, G. J., \& {Capak}, P. L. 2019, \apj, 882, 168

\bibitem[{{Pereira-Santaella} {et~al.}(2017){Pereira-Santaella}, {Rigopoulou},
  {Farrah}, {Lebouteiller}, \& {Li}}]{2017MNRAS.470.1218P}
{Pereira-Santaella}, M., {Rigopoulou}, D., {Farrah}, D., {Lebouteiller}, V., \& {Li}, J. 2017, \mnras, 470, 1218

\bibitem[{{Riechers} {et~al.}(2013){Riechers}, {Bradford}, {Clements},
  {Dowell}, {P{\'e}rez-Fournon}, {Ivison}, {Bridge}, {Conley}, {Fu}, {Vieira},
  {Wardlow}, {Calanog}, {Cooray}, {Hurley}, {Neri}, {Kamenetzky}, {Aguirre},
  {Altieri}, {Arumugam}, {Benford}, {B{\'e}thermin}, {Bock}, {Burgarella},
  {Cabrera-Lavers}, {Chapman}, {Cox}, {Dunlop}, {Earle}, {Farrah}, {Ferrero},
  {Franceschini}, {Gavazzi}, {Glenn}, {Solares}, {Gurwell}, {Halpern},
  {Hatziminaoglou}, {Hyde}, {Ibar}, {Kov{\'a}cs}, {Krips}, {Lupu}, {Maloney},
  {Martinez-Navajas}, {Matsuhara}, {Murphy}, {Naylor}, {Nguyen}, {Oliver},
  {Omont}, {Page}, {Petitpas}, {Rangwala}, {Roseboom}, {Scott}, {Smith},
  {Staguhn}, {Streblyanska}, {Thomson}, {Valtchanov}, {Viero}, {Wang},
  {Zemcov}, \& {Zmuidzinas}}]{2013Natur.496..329R}
{Riechers}, D.~A., {et~al.} 2013, \nat,
  496, 329

\bibitem[{{Rigopoulou} {et~al.}(2018){Rigopoulou}, {Pereira-Santaella},
  {Magdis}, {Cooray}, {Farrah}, {Marques-Chaves}, {Perez-Fournon}, \&
  {Riechers}}]{2018MNRAS.473...20R}
{Rigopoulou}, D., {Pereira-Santaella}, M., {Magdis}, G.~E., {Cooray}, A., {Farrah}, D., {Marques-Chaves}, R., {Perez-Fournon}, I., \& {Riechers}, D. 2018,
  \mnras, 473, 20

\bibitem[{{Schreiber} {et~al.}(2018){Schreiber}, {Glazebrook}, {Nanayakkara},
  {Kacprzak}, {Labb{\'e}}, {Oesch}, {Yuan}, {Tran}, {Papovich}, {Spitler}, \&
  {Straatman}}]{2018A&A...618A..85S}
{Schreiber}, C., {et~al.} 2018, \aap, 618,
  A85

\bibitem[{{Spilker} {et~al.}(2016){Spilker}, {Marrone}, {Aravena},
  {B{\'e}thermin}, {Bothwell}, {Carlstrom}, {Chapman}, {Crawford}, {de Breuck},
  {Fassnacht}, {Gonzalez}, {Greve}, {Hezaveh}, {Litke}, {Ma}, {Malkan},
  {Rotermund}, {Strandet}, {Vieira}, {Weiss}, \&
  {Welikala}}]{2016ApJ...826..112S}
{Spilker}, J.~S., {et~al.} 2016, \apj, 826,
  112

\bibitem[{{Strandet} {et~al.}(2016){Strandet}, {Weiss}, {Vieira}, {de Breuck},
  {Aguirre}, {Aravena}, {Ashby}, {B{\'e}thermin}, {Bradford}, {Carlstrom},
  {Chapman}, {Crawford}, {Everett}, {Fassnacht}, {Furstenau}, {Gonzalez},
  {Greve}, {Gullberg}, {Hezaveh}, {Kamenetzky}, {Litke}, {Ma}, {Malkan},
  {Marrone}, {Menten}, {Murphy}, {Nadolski}, {Rotermund}, {Spilker}, {Stark},
  \& {Welikala}}]{2016ApJ...822...80S}
{Strandet}, M.~L., {et~al.} 2016, \apj, 822, 80

\bibitem[{{Tadaki} {et~al.}(2019){Tadaki}, {Iono}, {Hatsukade}, {Kohno}, {Lee},
  {Matsuda}, {Michiyama}, {Nakanishi}, {Nagao}, {Saito}, {Tamura}, {Ueda}, \&
  {Umehata}}]{2019ApJ...876....1T}
{Tadaki}, K.-i., {et~al.} 2019, \apj, 876, 1

\bibitem[{{Tadaki} {et~al.}(2020){Tadaki}, {Belli}, {Burkert},
  {Dekel}, {F{\"o}rster Schreiber}, {Genzel}, {Hayashi}, {Herrera-Camus},
  {Kodama}, {Kohno}, {Koyama}, {Lee}, {Lutz}, {Mowla}, {Nelson}, {Renzini},
  {Suzuki}, {Tacconi}, {{\"U}bler}, {Wisnioski}, \&
  {Wuyts}}]{2020ApJ...901...74T}
{Tadaki}, K.-i., {et~al.} 2020, \apj,
  901, 74

\bibitem[{{Tamura} {et~al.}(2015){Tamura}, {Oguri}, {Iono}, {Hatsukade},
  {Matsuda}, \& {Hayashi}}]{2015PASJ...67...72T}
{Tamura}, Y., {Oguri}, M., {Iono}, D., {Hatsukade}, B., {Matsuda}, Y., \& {Hayashi}, M. 2015, \pasj, 67, 72

\bibitem[{{Tanaka} {et~al.}(2019){Tanaka}, {Valentino}, {Toft}, {Onodera},
  {Shimakawa}, {Ceverino}, {Faisst}, {Gallazzi}, {G{\'o}mez-Guijarro}, {Kubo},
  {Magdis}, {Steinhardt}, {Stockmann}, {Yabe}, \& {Zabl}}]{2019ApJ...885L..34T}
{Tanaka}, M., {et~al.} 2019, \apjl, 885, L34

\bibitem[{{Thomas} {et~al.}(2010){Thomas}, {Maraston}, {Schawinski}, {Sarzi},
  \& {Silk}}]{2010MNRAS.404.1775T}
{Thomas}, D., {Maraston}, C., {Schawinski}, K., {Sarzi}, M., \& {Silk}, J. 2010, \mnras, 404, 1775

\bibitem[{{Toft} {et~al.}(2014){Toft}, {Smol{\v{c}}i{\'c}}, {Magnelli},
  {Karim}, {Zirm}, {Michalowski}, {Capak}, {Sheth}, {Schawinski}, {Krogager},
  {Wuyts}, {Sanders}, {Man}, {Lutz}, {Staguhn}, {Berta}, {Mccracken}, {Krpan},
  \& {Riechers}}]{2014ApJ...782...68T}
{Toft}, S., {et~al.} 2014, \apj, 782,
  68

\bibitem[{{Torrey} {et~al.}(2019){Torrey}, {Vogelsberger}, {Marinacci},
  {Pakmor}, {Springel}, {Nelson}, {Naiman}, {Pillepich}, {Genel}, {Weinberger},
  \& {Hernquist}}]{2019MNRAS.484.5587T}
{Torrey}, P., {et~al.} 2019, \mnras, 484,
  5587

\bibitem[{{Valentino} {et~al.}(2020){Valentino}, {Tanaka}, {Davidzon}, {Toft},
  {G{\'o}mez-Guijarro}, {Stockmann}, {Onodera}, {Brammer}, {Ceverino},
  {Faisst}, {Gallazzi}, {Hayward}, {Ilbert}, {Kubo}, {Magdis}, {Selsing},
  {Shimakawa}, {Sparre}, {Steinhardt}, {Yabe}, \& {Zabl}}]{2020ApJ...889...93V}
{Valentino}, F., {et~al.} 2020, \apj, 889, 93

\bibitem[{{Wei{\ss}} {et~al.}(2013){Wei{\ss}}, {De Breuck}, {Marrone},
  {Vieira}, {Aguirre}, {Aird}, {Aravena}, {Ashby}, {Bayliss}, {Benson},
  {B{\'e}thermin}, {Biggs}, {Bleem}, {Bock}, {Bothwell}, {Bradford}, {Brodwin},
  {Carlstrom}, {Chang}, {Chapman}, {Crawford}, {Crites}, {de Haan}, {Dobbs},
  {Downes}, {Fassnacht}, {George}, {Gladders}, {Gonzalez}, {Greve},
  {Halverson}, {Hezaveh}, {High}, {Holder}, {Holzapfel}, {Hoover}, {Hrubes},
  {Husband}, {Keisler}, {Lee}, {Leitch}, {Lueker}, {Luong-Van}, {Malkan},
  {McIntyre}, {McMahon}, {Mehl}, {Menten}, {Meyer}, {Murphy}, {Padin},
  {Plagge}, {Reichardt}, {Rest}, {Rosenman}, {Ruel}, {Ruhl}, {Schaffer},
  {Shirokoff}, {Spilker}, {Stalder}, {Staniszewski}, {Stark}, {Story},
  {Vanderlinde}, {Welikala}, \& {Williamson}}]{2013ApJ...767...88W}
{Wei{\ss}}, A., {et~al.} 2013, \apj, 767, 88

\bibitem[{{Zavala} {et~al.}(2018){Zavala}, {Monta{\~n}a}, {Hughes}, {Yun},
  {Ivison}, {Valiante}, {Wilner}, {Spilker}, {Aretxaga}, {Eales},
  {Avila-Reese}, {Ch{\'a}vez}, {Cooray}, {Dannerbauer}, {Dunlop}, {Dunne},
  {G{\'o}mez-Ruiz}, {Micha{\l}owski}, {Narayanan}, {Nayyeri}, {Oteo}, {Rosa
  Gonz{\'a}lez}, {S{\'a}nchez-Arg{\"u}elles}, {Schloerb}, {Serjeant}, {Smith},
  {Terlevich}, {Vega}, {Villalba}, {van der Werf}, {Wilson}, \&
  {Zeballos}}]{2018NatAs...2...56Z}
{Zavala}, J.~A., {et~al.} 2018, Nature
  Astronomy, 2, 56

\end{thebibliography}

\end{document}